\documentclass[12pt]{article}
\usepackage{amssymb}
\usepackage{amsmath}

\makeatletter
\def\numberbysection{\@addtoreset{equation}{section}
 	\def\theequation{\thesection.\arabic{equation}}}
\makeatother

\numberbysection

\newcommand{\be}{\begin{eqnarray}}
\newcommand{\ee}{\end{eqnarray}}
\newcommand{\non}{\nonumber}
\newcommand{\tr}{\mathop{\rm tr}\nolimits}
\newcommand{\kk}{\kappa}
\newcommand{\id}{\mathbb{I}}
\newcommand{\csch}{\mathop{\rm csch}\nolimits}
\newcommand{\sech}{\mathop{\rm sech}\nolimits}
\newcommand{\M}{\mathop{\cal M}\nolimits}

\begin{document}

\begin{titlepage}
\strut\hfill UMTG--240
\vspace{.5in}
\begin{center}

\LARGE Bethe Ansatz solution of the open XXZ chain\\
\LARGE with nondiagonal boundary terms \\[1.0in]
\large Rafael I. Nepomechie\\[0.8in]
\large Physics Department, P.O. Box 248046, University of Miami\\[0.2in]  
\large Coral Gables, FL 33124 USA\\

\end{center}

\vspace{.5in}

\begin{abstract}
We propose a set of conventional Bethe Ansatz equations and a
corresponding expression for the eigenvalues of the transfer matrix
for the open spin-${1\over 2}$ XXZ quantum spin chain with nondiagonal
boundary terms, provided that the boundary parameters obey a certain
linear relation.
\end{abstract}
\end{titlepage}

\setcounter{footnote}{0}

\section{Introduction}\label{sec:intro}

Consider the open spin-${1\over 2}$ XXZ quantum spin chain with
nondiagonal boundary terms, defined by the Hamiltonian \cite{dVGR}
\be
{\cal H }&=& {1\over 2}\Big\{ \sum_{n=1}^{N-1}\left( 
\sigma_{n}^{x}\sigma_{n+1}^{x}+\sigma_{n}^{y}\sigma_{n+1}^{y}
+\cosh \eta\ \sigma_{n}^{z}\sigma_{n+1}^{z}\right)\non \\
&+&\sinh \eta \Big[ 
\coth \xi_{-} \sigma_{1}^{z}
+ {2 \kk_{-}\over \sinh \xi_{-}}\big( \cosh \theta_{-}\sigma_{1}^{x} 
+ i\sinh \theta_{-}\sigma_{1}^{y} \big) \non \\
&-& \coth \xi_{+} \sigma_{N}^{z}
- {2 \kk_{+}\over \sinh \xi_{+}}\big( \cosh \theta_{+}\sigma_{N}^{x}
+ i\sinh \theta_{+}\sigma_{N}^{y} \big)
\Big] \Big\} \,,
\label{Hamiltonian}
\ee
where $\sigma^{x} \,, \sigma^{y} \,, \sigma^{z}$ are the usual Pauli
matrices, $\eta$ is the bulk anisotropy parameter, $\xi_{\pm} \,,
\kk_{\pm} \,, \theta_{\pm}$ are arbitrary boundary parameters, and $N$
is the number of spins.  This is the prototypical integrable quantum 
spin chain with boundary. It is related to many other models, 
including the sine-Gordon field theory \cite{GZ}. Moreover, this model 
has applications in various branches of physics, including condensed 
matter and statistical mechanics.

This model has resisted solution for many years (see, e.g.,
\cite{Bat}).  The main difficulty is that, in contrast to the special
case of diagonal boundary terms (i.e., $\kk_{\pm}=0$) considered in
\cite{ABBBQ, Sk}, a simple pseudovacuum (reference) state does not
exist.  For example, the state with all spins up 
${1 \choose 0}^{\otimes N}$ is not an eigenstate
of the Hamiltonian.

We recently formulated \cite{JSP} a method of deriving the Bethe
Ansatz solution of integrable spin chain (vertex-type) models
which does not rely on the existence of a pseudovacuum state.  In
particular, we used this method to solve the model (\ref{Hamiltonian})
for the special case 
\be
\kk_{+} = \kk_{-} \,, \qquad 
\xi_{+} = \xi_{-} \,, \qquad 
\theta_{+} = \theta_{-} = 0\,, \qquad N =  \mbox{odd} \,.
\label{specialcase}
\ee

Here we propose the solution for a more general case.  Indeed, in
terms of the boundary parameters $\alpha_{\mp} \,, \beta_{\mp}$
introduced below in Eq.  (\ref{reparameter}), we find an expression
for the eigenvalues of the transfer matrix corresponding to the
Hamiltonian
\be
{\cal H }&=& {1\over 2}\Big\{ \sum_{n=1}^{N-1}\left( 
\sigma_{n}^{x}\sigma_{n+1}^{x}+\sigma_{n}^{y}\sigma_{n+1}^{y}
+\cosh \eta\ \sigma_{n}^{z}\sigma_{n+1}^{z}\right)\non \\
&+&\sinh \eta \Big[ 
\coth \alpha_{-} \tanh \beta_{-}\sigma_{1}^{z}
+ \csch \alpha_{-} \sech \beta_{-}\big( 
\cosh \theta_{-}\sigma_{1}^{x} 
+ i\sinh \theta_{-}\sigma_{1}^{y} \big) \non \\
&-& \coth \alpha_{+} \tanh \beta_{+} \sigma_{N}^{z}
+ \csch \alpha_{+} \sech \beta_{+}\big( 
\cosh \theta_{+}\sigma_{N}^{x}
+ i\sinh \theta_{+}\sigma_{N}^{y} \big)
\Big] \Big\} \,,
\label{newHamiltonian}
\ee
where the boundary parameters are subject to the linear relation
\be
\alpha_{-} + \beta_{-} + \alpha_{+} + \beta_{+} = \pm (\theta_{-} - 
\theta_{+}) + \eta k \,,
\label{constraintintro}
\ee
where $k$ is an even integer if $N$ is odd, and is an odd integer 
if $N$ is even.
In the recent paper \cite{CLSW}, similar results have been obtained 
by a different approach.

The outline of this article is as follows.  In Section
\ref{sec:transfer}, we briefly review the construction of the model's
transfer matrix, and list some of its important properties.  In
Section \ref{sec:BA}, we find the eigenvalues of the transfer matrix
by the three-step procedure formulated in \cite{JSP}.  The first two
steps, which lead to a functional relation for the transfer matrix,
are the same as in \cite{JSP, XXZ}, except for the introduction of the
parameters $\theta_{\mp}$.  The principal new results appear at
the third step, where we succeed to recast the functional relation
in terms of a determinant for the more general case
(\ref{constraintintro}).  We conclude with a brief discussion of our
results in Section \ref{sec:conclude}.

\section{The transfer matrix}\label{sec:transfer}

The fundamental transfer matrix $t(u)$
corresponding to the model (\ref{Hamiltonian}) is given by \cite{Sk}
\be
t(u) = \tr_{0} K^{+}_{0}(u)\  
T_{0}(u)\  K^{-}_{0}(u)\ \hat T_{0}(u)\,,
\label{transfer}
\ee
where the monodromy matrices are given by
\be
T_{0}(u) = R_{0N}(u) \cdots  R_{01}(u) \,,  \qquad 
\hat T_{0}(u) = R_{10}(u) \cdots  R_{N0}(u) \,, 
\label{monodromy}
\ee
and the $R$ matrix is the solution of the Yang-Baxter equation given
by
\be
R(u) = \left( \begin{array}{cccc}
	\sinh  (u + \eta) &0            &0           &0            \\
        0                 &\sinh  u     &\sinh \eta  &0            \\
	0                 &\sinh \eta   &\sinh  u    &0            \\
	0                 &0            &0           &\sinh  (u + \eta)
\end{array} \right) \,.
\label{bulkRmatrix}
\ee 
Moreover, the $K^{-}$ matrix is the solution of the boundary
Yang-Baxter equation \cite{Ch} given by \cite{dVGR, GZ}
\be
K^{-}(u) = \left( \begin{array}{cc}
\sinh(\xi_{-} + u)   & \kk_{-} e^{\theta_{-}} \sinh  2u \\
\kk_{-} e^{-\theta_{-}} \sinh  2u     & \sinh(\xi_{-} - u) 
\end{array} \right) \,,
\label{Kminusmatrix}
\ee 
which evidently depends on three boundary parameters $\xi_{-} \,, 
\kk_{-} \,, \theta_{-}$. It is related to the symmetric matrix 
$K^{-}(u)\Big\vert_{\theta_{-}=0}$ used in \cite{JSP, XXZ}
by a gauge transformation,
\be
K^{-}(u) = \M \ K^{-}(u)\Big\vert_{\theta_{-}=0}\ \M^{-1} \,, 
\label{gaugetransf}
\ee
with
\be
\M = \left( \begin{array}{cc}
e^{{1\over 2}\theta_{-}}  & 0\\
0                      & e^{-{1\over 2}\theta_{-}}
\end{array} \right) \,.
\label{gaugematrix}
\ee 
The matrix $K^{+}(u)$ is equal to $K^{-}(-u-\eta)$ with
$(\xi_{-} \,, \kk_{-} \,, \theta_{-})$ replaced by 
$(\xi_{+} \,, \kk_{+} \,, \theta_{+})$. Finally, 
$\tr_{0}$ denotes trace over the (two-dimensional)
``auxiliary space'' 0.  Further details about the construction of this
transfer matrix can be found in \cite{Sk, XXZ}.

The transfer matrix constitutes a one-parameter 
commutative family of matrices
\be
\left[ t(u)\,, t(v) \right] = 0  \,.
\label{commutativity}
\ee 
The Hamiltonian (\ref{Hamiltonian}) is related to the first derivative
of the transfer matrix,
\be
{\cal H} = c_{1} {\partial \over \partial u} t(u) \Big\vert_{u=0} 
+ c_{2} \id \,,
\label{firstderivative}
\ee
where
\be
c_{1} = {1\over 4 \sinh \xi_{-} \sinh \xi_{+} \sinh^{2N-1} \eta 
\cosh \eta} \,, \qquad 
c_{2} = - {\sinh^{2}\eta  + N \cosh^{2}\eta\over 2 \cosh \eta} 
\,,
\ee 
and $\id$ is the identity matrix.
The two relations (\ref{commutativity}),(\ref{firstderivative}) 
signal that the model is integrable. Moreover, it is evident that in 
order to determine the energy eigenvalues, it suffices to determine 
the eigenvalues of the transfer matrix.

The transfer matrix has the periodicity property
\be
t(u + i \pi)= t(u) \,,
\label{openperiodicity}
\ee
as well as crossing symmetry
\be
t(-u - \eta)= t(u) \,,
\label{transfercrossing}
\ee
and the asymptotic behavior (for $\kk_{\pm} \ne 0$)
\be
t(u) \sim -\kk_{-}\kk_{+} \cosh(\theta_{-}-\theta_{+})
{e^{u(2N+4)+\eta (N+2)}\over 2^{2N+1}} \id + 
\ldots \qquad \mbox{for} \qquad
u\rightarrow \infty \,.
\label{transfasympt}
\ee

\section{Bethe Ansatz solution}\label{sec:BA}

We now proceed to find an expression for the transfer matrix 
eigenvalues using the method formulated in \cite{JSP}. This method 
consists of three main steps:

\subsection{Step 1: fusion hierarchy}\label{step1}

The first step is to obtain the model's so-called fusion hierarchy. 
The transfer matrix (\ref{transfer}) is actually the first ($j={1\over
2}$) member of an infinite hierarchy of commuting transfer matrices
$t^{(j)}(u)$ corresponding to spin-$j$ (i.e., $(2j+1)$-dimensional)
auxiliary spaces, $j= {1\over 2}\,, 1\,, {3\over 2}\,, \ldots$.  Using
the fusion procedure for $R$ \cite{KS, KRS} and $K$ \cite{MN1, Zh}
matrices, one finds that these higher-level transfer matrices obey the
relations
\be
t^{(j)}(u) &=& \tilde \zeta_{2j-1}(2u+ (2j-1) \eta) \Big[
t^{(j-{1\over 2})}(u)\ t^{({1\over 2})}(u+ (2j-1)\eta ) \non \\
&-& {\Delta(u+(2j-2)\eta)\ \tilde \zeta_{2j-2}(2u+ (2j-2) \eta)  \over 
\zeta(2u+ 2(2j-1)\eta)}\ t^{(j-1)}(u) \Big] \,, 
\label{hierarchy}
\ee
with $t^{(0)} = \id$, and $j= 1 \,, {3\over 2}\,, \ldots $. 
The quantum determinant $\Delta (u)$ is given by
\be
\Delta (u) &=&
 -\left[\sinh(u + \eta + \xi_{-}) \sinh(u + \eta - \xi_{-})
+ \kk_{-}^{2} \sinh^{2}(2u+2\eta) \right] \non \\
&\times& \left[\sinh(u + \eta + \xi_{+}) \sinh(u + \eta - \xi_{+})
+ \kk_{+}^{2} \sinh^{2}(2u+2\eta) \right] \non \\
&\times& \sinh 2u \sinh(2u+4\eta)\ \zeta(u + \eta)^{2N} \,,
\label{qdeterminant}
\ee
and
\be
\tilde \zeta_{j}(u) &=& \prod_{k=1}^{j} \zeta(u + k \eta) \,, 
\qquad \tilde \zeta_{0}(u) = 1 \label{tildezeta} \,, \\
\zeta(u) &=& -\sinh (u+\eta) \sinh(u-\eta) \,.
\label{zeta}
\ee 

These relations are the same as those for the case of symmetric $K$
matrices ($\theta_{\mp}=0$) \cite{XXZ}.  We remark that the spin-$j$
matrix $K^{-}_{\langle 1 \ldots 2j \rangle}(u)$ is related to the
corresponding matrix with $\theta_{-}=0$ by a generalization of the
gauge transformation (\ref{gaugetransf}),
\be
K^{-}_{\langle 1 \ldots 2j \rangle}(u) =
\M_{1} \ldots \M_{2j}\
K^{-}_{\langle 1 \ldots 2j \rangle}(u)\Big\vert_{\theta_{-}=0}\
\M_{2j}^{-1} \ldots \M_{1}^{-1} \,.
\label{spinjgaugetransf}
\ee 

\subsection{Step 2: truncation at roots of unity}\label{step2}

The second step is to observe \cite{XXZ} that for anisotropy values
\be
\eta = {i \pi\over p+1}\,, \qquad p= 1 \,, 2 \,, \ldots \,,
\label{etavalues}
\ee
(and hence $q \equiv e^{\eta}$ is a root of
unity, satisfying $q^{p+1}=-1$), the level-${p+1\over 2}$ transfer
matrix can be expressed in terms of a transfer matrix of one level
lower,
\be
t^{({p+1\over 2})}(u) 
= \alpha(u) \left[ 
t^{({p-1\over 2})}(u+ \eta)
+ \beta(u) \id \right]
\,. \label{truncation}
\ee
The quantities $\alpha(u)$ and $\beta(u)$ are given by the
corresponding expressions (4.31) in \cite{XXZ}, except with
$\sigma_{\mp}(u) \rightarrow e^{(p+1)\theta_{\mp}} \sigma_{\mp}(u)$
and $\rho_{\mp}(u) \rightarrow e^{-(p+1)\theta_{\mp}}\rho_{\mp}(u)$,
as a consequence of (\ref{spinjgaugetransf}). 

This result provides an example of McCoy's dictum ``Complicated is
simple'' \cite{Mc}.  Indeed, the essential point of this step is to
exploit the higher symmetry which occurs at roots of unity to help
solve the model.

Combining the fusion hierarchy (\ref{hierarchy}) and the truncation
identity (or ``closing relation'') (\ref{truncation}) for the $\eta$ values 
(\ref{etavalues}), we arrive at a functional relation for the 
fundamental transfer matrix $t(u) \equiv t^{({1\over 2})}(u)$ (and hence, 
for the corresponding eigenvalues $\Lambda(u)$) of order $p+1$ 
\cite{JSP, XXZ}:
\be
\lefteqn{\Lambda(u) \Lambda(u +\eta) \ldots \Lambda(u + p \eta)} \non \\
&-& \delta (u-\eta) \Lambda(u +\eta) \Lambda(u +2\eta) 
\ldots \Lambda(u + (p-1)\eta) \non \\
&-& \delta (u) \Lambda(u +2\eta) \Lambda(u +3\eta)
\ldots \Lambda(u + p \eta) \non \\
&-& \delta (u+\eta) \Lambda(u) \Lambda(u +3\eta) \Lambda(u +4\eta) 
\ldots \Lambda(u + p \eta) \non \\
&-& \delta (u+2\eta) \Lambda(u) \Lambda(u +\eta) \Lambda(u +4\eta) 
\ldots \Lambda(u + p \eta) - \ldots \non \\
&-& \delta (u+(p-1)\eta) \Lambda(u) \Lambda(u +\eta) 
\ldots \Lambda(u +  (p-2)\eta) \non \\
&+& \ldots  = f(u) \,,
\label{funcrltn}
\ee 
where $\delta(u)$ is defined by
\be
\delta(u) = {\Delta (u)\over \zeta(2u+2\eta)} \,.
\label{delta}
\ee
Moreover, the 
function $f(u)$ is given by 
\be
f(u) &=& {(-1)^{p (N+1)}\over 2^{2p(N+1)}} \sinh^{2N}((p+1)u) 
{\cosh^{2}((p+1)u +{i \pi\over 2}\epsilon)\over \cosh^{2}((p+1)u)}
\non \\
&\times& \Big\{ 
n(u \,; \xi_{-} \,, \kk_{-})\ n(u \,; -\xi_{+} \,, \kk_{+}) +
n(u \,; -\xi_{-} \,, \kk_{-})\ n(u \,; \xi_{+} \,, \kk_{+}) \non \\
&\quad&+ 2(-1)^{N} (-\kk_{-} \kk_{+})^{p+1} \sinh^{2}(2(p+1)u) 
\cosh((p+1)(\theta_{-} - \theta_{+})) \Big\} \,,
\label{ffunction}
\ee
where $\epsilon= 2 \mbox{frac}(p/2)$ equals 0 if $p$ is even, and
equals 1 if $p$ is odd; and the function $n(u \,; \xi \,, \kk)$ is
defined by
\be
n(u \,; \xi \,, \kk) = \sinh \left( (p+1)(\xi +u) \right)  
+ \sum_{l=1}^{\left[{p+1\over 2}\right]}c_{p\,, l}\ 
\kk^{2l} \sinh \left( (p+1)u + (p+1 - 2l) \xi \right) \,,
\label{nfunction}
\ee 
with
\be
c_{p \,, l} = {(p+1)\over l!} \prod_{k=0}^{l-2} (p-l-k) 
\,. \non 
\ee

For instance, for the case $p=3$, the functional relation is given 
by 
\be
\Lambda(u) \Lambda(u +\eta) \Lambda(u +2\eta) \Lambda(u +3\eta) 
- \delta(u-\eta) \Lambda(u +\eta) \Lambda(u +2\eta) 
\qquad \qquad \qquad \qquad \non  \\
 - \delta(u) \Lambda(u +2\eta) \Lambda(u +3\eta) 
- \delta(u +\eta) \Lambda(u) \Lambda(u +3\eta)   
- \delta(u +2\eta) \Lambda(u) \Lambda(u +\eta)  \non  \\ 
+ \delta(u) \delta(u +2\eta)
+ \delta(u-\eta) \delta(u +\eta) = f(u) \,.
\ee 
 
\subsection{Step 3: determinant representation}\label{step3}

Following the strategy used in \cite{BR} to solve RSOS models, the
third and final step is to rewrite the functional relation as the
determinant of a $(p+1) \times (p+1)$ matrix.  Let us assume that this
matrix has the same form as the one for the diagonal case
($\kk_{\pm}=0$) and for the case (\ref{specialcase}).  That is, we
assume the functional relation can be cast in the form \cite{JSP}
\be
\det \left(
\begin{array}{cccccccc}
    \Lambda_{0} & -h'_{-1} & 0 & 0 & \ldots & 0 & 0 & -h_{0}  \\
    -h_{1} & \Lambda_{1} & -h'_{0} & 0 & \ldots & 0 & 0 & 0  \\
    0 & -h_{2} & \Lambda_{2} & -h'_{1} & \ldots & 0 & 0 & 0  \\
    \vdots  & \vdots & \vdots & \vdots & \ddots 
    & \vdots  & \vdots & \vdots   \\
    0 & 0 & 0 & 0 & \ldots & -h_{p-1} & \Lambda_{p-1} & -h'_{p-2}  \\
    -h'_{p-1} & 0 & 0 & 0 & \ldots & 0 & -h_{p} & \Lambda_{p}
\end{array} \right) = 0 \,,
\label{opendetform}
\ee
where $\Lambda_{k} = \Lambda(u + \eta k)$, $h_{k} = h(u + \eta k)$, 
$h'_{k} = h'(u + \eta k)$, 
\be 
h'(u) = h(-u-2\eta) \,, 
\label{functionhprime}
\ee 
and the function $h(u)$ is yet to be determined.  We find that the
functional relation (\ref{funcrltn}) can indeed be recast in the form
(\ref{opendetform}), provided that $h(u)$ satisfies the three
conditions
\be
h(u + i \pi) &=& h(u) \label{cond0} \\
h(u+\eta) h(-u-\eta) &=& \delta(u) \,, \label{cond1} \\
\prod_{j=0}^{p} h(u+j\eta) + \prod_{j=0}^{p} h(-u-j\eta) &=& f(u) 
\,. \label{cond2} 
\ee
The results \cite{JSP} for $h(u)$ in the diagonal case and in the case
(\ref{specialcase}) suggest that, in general, $h(u)$ has form
\be 
h(u) = -\sinh^{2N}(u+\eta){\sinh(2u+2\eta)\over \sinh(2u+\eta)}
g_{-}(u) g_{+}(u) \,, \label{hAnsatz} 
\ee
where the functions $g_{\mp}(u)$ contain all the dependence on the
boundary parameters.

Then second condition (\ref{cond1}) together with
(\ref{qdeterminant}), (\ref{delta}) and (\ref{hAnsatz}) imply that
\be
g_{-}(u) g_{+}(u) g_{-}(-u) g_{+}(-u) &=& 
(\sinh^{2}u -\sinh^{2} \xi_{-} 
+ \kk_{-}^{2} \sinh^{2}2u ) \non \\
&\times& (\sinh^{2}u -\sinh^{2} \xi_{+} 
+ \kk_{+}^{2} \sinh^{2}2u )
\,. 
\label{gconstraint}
\ee
This suggests that $g_{\mp}(u)$ obey the functional equation
\be
g_{\mp}(u) g_{\mp}(-u) = -(\sinh^{2}u -\sinh^{2} \xi_{\mp} 
+ \kk_{\mp}^{2} \sinh^{2}2u ) \,.
\label{gfunceq}
\ee
Assuming that the functions $g_{\mp}(u)$ are given by
\be
g_{\mp}(u) = 2 \kk_{\mp} \sinh(u + \alpha_{\mp}) \cosh(u + \beta_{\mp}) 
\,,
\label{gmp}
\ee
then (\ref{gfunceq}) implies that the parameters $\alpha_{\mp}$, 
$\beta_{\mp}$ obey
\be
\sinh^{2} \alpha_{\mp} \cosh^{2} \beta_{\mp} 
= {1\over 4\kk_{\mp}^{2}} \sinh^{2} \xi_{\mp} \,, 
\qquad
\cosh^{2}\alpha_{\mp} \sinh^{2}\beta_{\mp}
= {1\over 4\kk_{\mp}^{2}} \cosh^{2} \xi_{\mp} \,.
\label{reparametersquare}
\ee
A similar reparametrization appears in \cite{GZ, CLSW}.  Below we
shall argue that (\ref{gmp}) is essentially the unique solution of
(\ref{gconstraint}).

The third condition (\ref{cond2}) together with (\ref{ffunction}) and
(\ref{hAnsatz}) imply that \footnote{Note that the right-hand-side 
of (\ref{gprod}) depends on $N$ only through its parity $(-1)^{N}$.}
\be
& &\prod_{j=0}^{p} g_{-}(u+j\eta) g_{+}(u+j\eta) + 
\prod_{j=0}^{p} g_{-}(-u-j\eta) g_{+}(-u-j\eta) \non \\ 
&=& {(-1)^{p}\over 2^{2p}} \Big\{ 
n(u \,; \xi_{-} \,, \kk_{-})\ n(u \,; -\xi_{+} \,, \kk_{+}) +
n(u \,; -\xi_{-} \,, \kk_{-})\ n(u \,; \xi_{+} \,, \kk_{+}) \non \\
&\quad&+ 2(-1)^{N} (-\kk_{-} \kk_{+})^{p+1} \sinh^{2}(2(p+1)u) 
\cosh((p+1)(\theta_{-} - \theta_{+})) \Big\} \,,
\label{gprod}
\ee
where $n(u \,; \xi \,, \kk)$ is given by Eq. (\ref{nfunction}).
We find that this requirement can be satisfied for 
$p=\mbox{odd}$, with  $g_{\mp}(u)$ given by (\ref{gmp}) 
and \footnote{The requirement of including the special case
(\ref{specialcase}), which corresponds to 
$\alpha_{-}=-\alpha_{+}$, $\beta_{-}=-\beta_{+}$, $k=0$, helps to
resolve the sign ambiguity in passing from (\ref{reparametersquare})
to (\ref{reparameter}).}
\be
\sinh \alpha_{-} \cosh \beta_{-} &=& {1\over 2\kk_{-}} \sinh \xi_{-} \,, 
\qquad
\cosh \alpha_{-} \sinh \beta_{-} = {1\over 2\kk_{-}} \cosh \xi_{-} \,, 
\non \\
\sinh \alpha_{+} \cosh \beta_{+} &=& -{1\over 2\kk_{+}} \sinh \xi_{+} \,, 
\qquad
\cosh \alpha_{+} \sinh \beta_{+} = -{1\over 2\kk_{+}} \cosh \xi_{+} \,, 
\label{reparameter}
\ee
provided that the various parameters obey the linear constraint
\be
\alpha_{-} + \beta_{-} + \alpha_{+} + \beta_{+} = \pm (\theta_{-} - 
\theta_{+}) + \eta k \,,
\label{constraint}
\ee
where $k$ is an even integer if $N$ is odd, and is an odd integer 
if $N$ is even.

In short, the functional relations can be cast in the determinant 
form (\ref{opendetform}) for $p=\mbox{odd}$ with
\be 
h(u) &=& -\sinh^{2N}(u+\eta){\sinh(2u+2\eta)\over \sinh(2u+\eta)} 
\non \\
&\times& 4 \kk_{-} \kk_{+} \sinh(u + \alpha_{-}) \cosh(u + \beta_{-}) 
\sinh(u + \alpha_{+}) \cosh(u + \beta_{+}) 
\,, \label{hfinal} 
\ee
where $\alpha_{\mp}$, $\beta_{\mp}$ are defined by (\ref{reparameter})
and satisfy the constraint (\ref{constraint}).

We now proceed as in \cite{BR, JSP}, and assume that the matrix
in (\ref{opendetform}) has the null vector
$(Q_{0} \,, Q_{1} \,, \ldots \,, Q_{p})$. That is,
\be
\Lambda_{0} Q_{0} - h'_{-1} Q_{1} - h_{0} Q_{p} &=& 0 \,,  
\non \\
-h_{k} Q_{k-1} + \Lambda_{k} Q_{k} - h'_{k-1} Q_{k+1} &=& 0 \,, \qquad
k = 1\,, \ldots \,, p-1  \,, \non \\
-h'_{p-1} Q_{0} - h_{p} Q_{p-1} + \Lambda_{p} Q_{p}  &=& 0 \,.
\label{nullopen}
\ee
We make the Ansatz $Q_{k} = Q(u + \eta k)$, where $Q(u)$ is given by
\be
Q(u) = \prod_{j=1}^{M} \sinh(u - u_{j}) \sinh(u + u_{j} + \eta) \,,
\label{openQ}
\ee 
which has the crossing symmetry $Q(u) = Q(-u-\eta)$. The zeros $u_{j}$
of $Q(u)$ are still to be determined.
Eqs. (\ref{nullopen}) and (\ref{functionhprime}) imply
that the eigenvalues are given by
\be
\Lambda(u) = h(u) {Q(u-\eta)\over Q(u)} 
+ h(-u-\eta) {Q(u+\eta)\over Q(u)}  \,.
\label{openeigenvalues} 
\ee
We verify that this result is consistent with both
the periodicity (\ref{openperiodicity}) and crossing 
(\ref{transfercrossing}) properties of the transfer matrix.
The requirement that $\Lambda(u)$ be analytic at $u=u_{j}$ 
yields the Bethe Ansatz equations
\be
{h(u_{j})\over h(-u_{j}-\eta)} = 
-{Q(u_{j}+\eta)\over Q(u_{j}-\eta)} \,, 
\qquad j = 1 \,, \ldots \,, M \,.
\label{openBAeqs}
\ee
The asymptotic behavior (\ref{transfasympt}), together with the result
(\ref{openeigenvalues}) for the eigenvalues and the constraint
(\ref{constraint}), imply that the number $M$ of Bethe roots is given
by
\be
M={1\over 2}(N-1+k) \,,
\label{Mvalue}
\ee
where $k$ is the integer appearing in (\ref{constraint}).
We leave to a future investigation the interesting question of
determining further restrictions on the value of $k$, which presumably
is related to the question of completeness.

We now argue that (\ref{gmp}) is essentially the unique solution of 
(\ref{gconstraint}). 
Indeed, if $\tilde g_{\mp}(u)$ are also solutions of 
(\ref{gconstraint}), then 
\be
\tilde g_{-}(u) \tilde g_{+}(u) = g_{-}(u) g_{+}(u) \phi(u) \,,
\ee
where $g_{\mp}(u)$ are given by (\ref{gmp}), and $\phi(u)$ satisfies
\be
\phi(u) \phi(-u) = 1 \,.
\label{phi1}
\ee
The periodicity condition (\ref{cond0}) implies that $\phi(u)$
has the same periodicity
\be
\phi(u + i \pi) = \phi(u) \,.
\label{phi2}
\ee
We infer from (\ref{phi1}) and (\ref{phi2}) that $\phi(u)$ is a
CDD-like factor
\be
\phi(u) = \prod_{j} {\sinh(u + v_{j})\over \sinh(u - v_{j})} \,.
\ee
The requirement that $\Lambda(u)$ be analytic then restricts $\phi(u)$ 
to the form
\be
\phi(u) = {q(u - \eta)\over q(u)} \,, \qquad \mbox{where} \qquad 
q(u) = \prod_{j} \sinh(u - v_{j}) \sinh(u + \eta + v_{j}) \,.
\ee
This is equivalent to having additional Bethe roots, which can be 
included in $Q(u)$ (\ref{openQ}).

Although the above results for the eigenvalues (\ref{hfinal}),
(\ref{openeigenvalues}), (\ref{openBAeqs}) have been obtained under the
assumption that $\eta$ is restricted to the values (\ref{etavalues})
with $p$ odd, we expect that these results remain valid for generic
values of $\eta$.  Indeed, we have explicitly verified that these
expressions reproduce the correct eigenvalues for $N=0$ (with $M=0$)
and $N=1$ (with $M=1$) for arbitrary $\eta$.

\section{Conclusion}\label{sec:conclude}

Our proposed expression for the eigenvalues $\Lambda(u)$ of the
transfer matrix (\ref{transfer}) corresponding to the Hamiltonian
(\ref{newHamiltonian}) is given by (\ref{openeigenvalues}), where
$h(u)$ and $Q(u)$ are given by (\ref{hfinal}) and (\ref{openQ}),
respectively; the Bethe Ansatz equations are given by
(\ref{openBAeqs}), with $M$ given by (\ref{Mvalue}); and the
parameters $\alpha_{\mp}$, $\beta_{\mp}$ (which are related to
$\xi_{\mp}$, $\kk_{\mp}$ by (\ref{reparameter})) must satisfy the
constraint (\ref{constraint}).

It remains an open question whether a solution with Bethe Ansatz
equations of the ``conventional'' form (\ref{openBAeqs}) can be found
which does not require a constraint among the boundary parameters. 
(Although the solution proposed in \cite{XXZ} does not require any
constraint among the boundary parameters, it holds only for the $\eta$
values (\ref{etavalues}), and the Bethe Ansatz equations are not of
the conventional form.)

For the special case (\ref{specialcase}), an analysis of the
thermodynamic ($N \rightarrow \infty$) limit and an extension to
higher-dimensional representations has recently been given in
\cite{Do}.  For the more general case discussed here, it should now be
possible to address such questions, and also to find generalizations
to higher rank-algebras, both for the trigonometric and elliptic
cases.

\section*{Acknowledgments}

I am grateful to all the organizers of the Annecy workshop ``Recent
Advances in the Theory of Quantum Integrable Systems'' for the
opportunity to present this work.  I am also grateful to many of the
participants, in particular A. Belavin, A. Doikou, B. McCoy, F.
Ravanini, V. Rittenberg and S. Ruijsenaars, for their questions or
comments.  This work was supported in part by the National Science
Foundation under Grant PHY-0098088.


\begin{thebibliography}{99}

\bibitem{dVGR}
H.J. de Vega and A. Gonz\'alez-Ruiz, J. Phys. {\it A26} (1993) L519.

\bibitem{GZ}
S. Ghoshal and A.B. Zamolodchikov, Int. J. Mod. Phys. {\it A9} (1994)
3841.

\bibitem{Bat}
M.T. Batchelor, in Proceedings of the 22nd International Colloquium on
Group Theoretical Methods in Physics, eds S.P. Corney {\it et al.},
International Press, Boston (1999)  261.

\bibitem{ABBBQ}
F.C. Alcaraz, M.N. Barber, M.T. Batchelor, R.J. Baxter and G.R.W. Quispel,
J. Phys. {\it A20} (1987) 6397. 

\bibitem{Sk}
E.K. Sklyanin, J. Phys. {\it A21} (1988) 2375.

\bibitem{JSP}
R.I. Nepomechie, J. Stat. Phys. {\it 111} (2003) 1363.

\bibitem{CLSW}
J. Cao, H.-Q. Lin, K.-J. Shi, Y. Wang, 
``Exact solutions and elementary excitations in the XXZ spin chain with 
unparallel boundary fields,''  {\tt cond-mat/0212163}.

\bibitem{XXZ} 
R.I. Nepomechie, Nucl. Phys. {\it B622} (2002) 615; Addendum, 
Nucl. Phys. {\it B631} (2002) 519.

\bibitem{Ch} 
I.V. Cherednik, Theor. Math. Phys. {\it 61} (1984) 977.

\bibitem{KS}
P.P. Kulish and E.K. Sklyanin, {\it Lecture Notes in Physics}, Vol. 151,
(Springer, 1982) 61.

\bibitem{KRS}
P.P. Kulish, N.Yu. Reshetikhin and E.K. Sklyanin, 
Lett. Math. Phys. {\it 5} (1981) 393.

\bibitem{MN1}
L. Mezincescu and R.I. Nepomechie, J. Phys. {\it A25} (1992) 2533.

\bibitem{Zh}
Y.-K. Zhou, Nucl. Phys. {\it B458} (1996) 504.

\bibitem{Mc}
B. McCoy, talk at the Annecy workshop ``Recent
Advances in the Theory of Quantum Integrable Systems, '' March 2003.

\bibitem{BR} 
V.V. Bazhanov and N.Yu.  Reshetikhin, Int.  J. Mod.  Phys. 
{\it A4} (1989) 115.

\bibitem{Do}
A. Doikou, ``Fused integrable lattice models with quantum impurities and 
open boundaries,'' {\tt hep-th/0303205}.



\end{thebibliography}
\end{document}